\documentclass[aps,prd,twocolumn,groupedaddress]{revtex4-1}
\usepackage{bbm,amsthm,amsmath,amssymb,cancel,mathrsfs,graphics,multirow,array,enumerate}

\newcommand{\mbf}[1][k]{\mathbf{#1}}
\newcommand{\cc}[1][k]{\text{c.c.}}

\begin{document}

\preprint{TCC-007-11}

\title{Calculating the local-type $f_{\text{NL}}$ for slow-roll inflation with a non-vacuum initial state}

\author{Jonathan Ganc}
\email{jonganc@physics.utexas.edu}
\affiliation{Texas Cosmology Center and Department of Physics, The University of Texas at Austin, Austin, TX 78712, USA}

\date{\today}

\begin{abstract}
Single-field slow-roll inflation with a non-vacuum initial state has an enhanced bispectrum in the local limit. We numerically calculate the local-type $f_{\text{NL}}$ signal in the CMB that would be measured for such models (including the full transfer function and 2D projection).
The nature of the result depends on several parameters, including the occupation number $N_k$, the phase angle $\theta_k$ between the Bogoliubov parameters, and the slow-roll parameter $\epsilon$. In the most conservative case, where one takes $\theta_k \approx \eta_0 k$ (justified by physical reasons discussed within) and $\epsilon\lesssim 0.01$, we find that $0 < f_{\text{NL}} < 1.52 (\epsilon/0.01)$, which is likely too small to be detected in the CMB. However, if one is willing to allow a constant value for the phase angle $\theta_k$ and $N_k=\mathcal{O}(1)$, $f_{\text{NL}}$ can be much larger and/or negative (depending on the choice of $\theta_k$), e.g. $f_{\text{NL}} \approx 28 (\epsilon/0.01)$ or $-6.4 (\epsilon/0.01)$; depending on $\epsilon$, these scenarios could be detected by Planck or a future satellite. While we show that these results are not actually a violation of the single-field consistency relation, they do produce a value for $f_{\text{NL}}$ that is considerably larger than that usually predicted from single-field inflation.
\end{abstract}


\maketitle

\section{Introduction}
\label{sec:intro}
Most cosmologists share the belief that the universe underwent an early inflationary period, where it experienced an accelerated expansion \cite{Guth:1980zm,*Linde:1981mu,*Albrecht:1982wi} that produced the large-scale perturbations that seeded the cosmic microwave background (CMB) anisotropy and large-scale structure \cite{Guth:1982ec,*Starobinsky:1982ee,*Hawking:1982cz,*Bardeen:1983qw,*Mukhanov:1981xt}. However, understanding the mechanism and exact history of this early period is still the subject of active research and the impetus behind the development of a variety of inflationary models \cite{*[{For a review, see, e.g., }] [{}] Liddle:2000cg}. One of the most promising avenues for discrimination between different models is the measurement of the non-Gaussianity of the primordial curvature perturbation $\zeta$, that is, the degree to which the primordial fluctuations contain statistical information beyond the power spectrum \cite{*[{For a review, see, e.g., }] [{}] Bartolo:2004if}. As it stands, all measurements are consistent with a Gaussian curvature perturbation \cite{Komatsu:2010fb}; however, this analysis may change with the release of data from the Planck satellite, as well as large-scale structure observations.

The bispectrum is the lowest-order non-Gaussianity and, at least in most models, the most likely to be detected. It is defined in terms of the Fourier transform of the three-point function
\begin{align}
  \left\langle \zeta_{\mbf_1} \zeta_{\mbf_2} \zeta_{\mbf_3} \right\rangle
  = (2 \pi)^3 \delta^{(3)}\!\big({\textstyle\sum} \mbf_i\big) \,
    B_\zeta(k_1,k_2,k_3)\,,
\end{align}
where the $\delta$-function is required by translation invariance of the underlying theory. The bispectrum $B_\zeta(k_1, k_2, k_3)$ is a real-valued function of three variables; however, the measured data has only limited signal strength so the bispectrum is often reported in terms of $f_{\text{NL}}$ parameters, each of which is obtained by fitting the measurements to a template bispectrum. This paper will deal with the local bispectrum $f_{\text{NL}}^{\text{loc}}$ (henceforth simply called $f_{\text{NL}}$), the most commonly reported non-Gaussianity parameter. The local template is the result of assuming that the non-Gaussianity is produced from a background Gaussian field $\zeta_g$ by a purely-local process \cite{Komatsu:2001rj}, so that 
\begin{align}\label{eq:loc_ans}
  \zeta(\mbf[x]) = \zeta_g(\mbf[x])
         + (3/5) f_{\text{NL}} \zeta_g^2(\mbf[x]) + \ldots\; ;
\end{align}
a straightforward calculation then yields the bispectrum
\begin{align}\label{eq:loc_bispr}
  B_\zeta^{\text{loc}}(k_1,k_2,k_3)
    = \frac{6}{5} f_{\text{NL}}
     [P_\zeta(k_1) & P_\zeta(k_2) + P_\zeta(k_2) P_\zeta(k_3)+\cr
    &+ P_\zeta(k_3) P_\zeta(k_1)] \,.
\end{align}
Noting that the measured power spectrum $P_\zeta(k) \propto k^{-3}$ and that the wavenumbers must obey $\mbf[k]_1+\mbf[k]_2+\mbf[k]_3=0$, we see that this bispectrum peaks when one of the wavenumbers is much smaller than the other two, e.g. $k_3 \ll k_1 \approx k_2$; this limit is known as the \emph{squeezed-limit} or \emph{local-limit} \cite{Babich:2004gb}. In this regime, the bispectrum becomes
\begin{align}\label{eq:loc_bispr_sq_lim}
  B_\zeta^{\text{loc}}(k_1,k_2,k_3)_{k_3\ll k_1\approx k_2}
   \approx\frac{12}{5}f_{\text{NL}}P(k_1)P(k_3).
\end{align}
$f_{\text{NL}}$ gained much importance when Creminelli and Zaldarriaga \cite{creminelli04} demonstrated a consistency relation that shows that for all single-field inflation models (regardless of kinetic term, vacuum state, etc.) in the squeezed-limit
\begin{align*}
  B_\zeta (k_3\ll k_1)=(1-n_s) P(k_1) P(k_3).
\end{align*}
Since the local template is most sensitive to squeezed-limit wavenumber configurations, the consistency relation implies that 
\begin{align*}
  f_{\text{NL}}\approx(5/12) (1-n_s)\approx 0.01,
\end{align*}
where $n_s$ is the spectral tilt; thus, a larger detection of $f_{\text{NL}}$ would strongly disfavor single-field inflation. The consistency relation has been explored in many papers, including \cite{Malda,seery_05sf,*Cheung:2007sv,*RenauxPetel:2010ty,*Das:2009sg,*Alberghi:2010sn,Chen06,Ganc:2010ff,*[{Interestingly, it can be hard to produce large $f_{\text{NL}}$ even in multi-field models; see }] [{}] Meyers:2010rg}.
However, one must be cautious when using the consistency relation to interpret observations. The relation is inviolable only in the exact squeezed limit while measurements are made over a finite range of $k$ values and involve a best-fit of the measured bispectrum to the local-template bispectrum. Recently, Agullo and Parker \cite{Agullo:2010ws} showed that there can be an enhancement of the local-limit bispectrum for slow-roll inflation given a non-vacuum initial state. In \S~\ref{sec:swr-inf}, we reproduce their result, though in a formalism where the non-standard initial state is recorded in the Bogoliubov parameters (rather than by a density matrix). We then proceed, in \S~\ref{sec:obs_fnl} to calculate the value of $f_{\text{NL}}$ that such a model would produce in the CMB. In \S~\ref{sec:results} we report the results, which we interpret further in \S~\ref{sec:discuss}. We find that $f_{\text{NL}} \lesssim 3$ (i.e. undetectable) in the most conservative case but that we can have $f_{\text{NL}}\gg1$ if we allow some freedom in the parameters; this result highlights the importance of properly applying the consistency relation.

\section{Slow-roll inflation with a non-Bunch-Davies initial state}
\label{sec:swr-inf}

First, we must calculate the bispectrum for slow-roll inflation with a non-vacuum initial state. The fluctuations in the scalar curvature perturbation $\zeta$ are produced by quantum fluctuations in a scalar field $\phi$ which has the action 
\begin{align*}
  S = \frac{1}{2} \int d^4x \sqrt{-g } [R - (\nabla \phi)^2 - 2 V(\phi)]\,.
\end{align*}
We suppose that the slow-roll parameters are small, i.e. 
\begin{align*}
  \epsilon &\equiv \frac{1}{2} \dot \phi^2 / H^2 \ll 1\cr
  \eta_\text{SR} &\equiv - \ddot \phi / H \dot \phi + (1/2)\, \dot
  \phi^2/H^2\ll 1.
\end{align*}
 After switching the degree-of-freedom to $\zeta$, we can calculate the action to third order (note that $M_{\text{pl}}^{-2}\equiv8\pi G\equiv1$) as \cite{Malda}
\begin{align}
 \label{eq:acn}
 S_2 = & \frac{1}{2} \int d^4x \frac{\dot \phi^2}{H^2} 
  [a^3 \dot \zeta^2 - a (\partial \zeta)^2],\cr
  S_3 = & \int d^4x \frac{\dot \phi^4}{H^4} a^5 H \dot \zeta^2 \partial^{-2} \dot \zeta
    \equiv \int dt \,L_3(t),
\end{align}
where as usual $a$ is the scale factor and $H\equiv\dot a/a$; note that we have neglected a field redefinition which is a small effect and irrelevant for this discussion. Then, to lowest order in slow-roll, the equation of motion for $\zeta$ is
\begin{align}
  \label{eq:eom}
  - \partial_t ( a^{ 3} \frac{ \dot \phi_0^2}{H^2} \dot \zeta ) + a
  \frac{ \dot \phi_0^2}{H^2} \partial^2 \zeta=0.
\end{align}

Using the in-in formalism, the tree-level three-point function for $\zeta$ is \cite{weinberg_qccc}
\begin{align}\label{eq:in-in}
  \big\langle  \zeta^3(t^*) \big\rangle
   = - i \int_{t_0}^{t^*} dt' 
    \big\langle 0\big| [\zeta^3(t^*), H_I(t')]\big| 0\big\rangle\,,
\end{align}
where the brackets on the LHS denote the ensemble average, $H_I$ is the interaction Hamiltonian (and equals $-L_3$), and $t_0$ is the time where we set initial conditions on the state of the universe. To evaluate the RHS, we quantize $\zeta$ using Quantum Field Theory in curved space-time \cite{*[{See, e.g., }] [{}] birell&davies}:
\begin{align}\label{eq:qntz_zet}
  \zeta_{\mbf[k]}(t) = u_k(t) a_{\mbf} + u^*_k(t) a^\dagger_{-\mbf}\,,
\end{align}
where $a_{\mbf[k]}, a^\dagger_{\mbf[k]}$ are annihilation, creation operators, respectively, for $\zeta$ modes and $u_k(t)$ are the mode functions for $\zeta$, i.e. complex solutions to the equations of motion \eqref{eq:eom} satisfying some normalization condition. We usually choose $u$ to correspond to the so-called Bunch-Davies (BD) vacuum, where it contains only positive frequency modes and where the vacuum state is unchanged under transformations that leave the metric de-Sitter; for slow-roll inflation, the BD vacuum state is
\begin{align}\label{eq:BD_mode_fcn}
  u_k(\eta)=  \frac{H^2}{\dot \phi} \frac{1}{\sqrt{2 k^3}}
    (1+ik \eta) e^{-ik \eta}.
\end{align}

More recently, people have realized that the assumption of a BD vacuum may be too restrictive and have considered initial states containing particles \cite{Martin:1999fa,*Chen09,Boyanovsky:2006qi,Chen06,Holman:2007na,Meerburg:2009ys,Meerburg:2009fi,Agullo:2010ws}. These states can be thought of as parameterizing either early-time or high-energy physics and can be represented by performing a Bogoliubov transformation on the BD $u$:
\begin{align}\label{eq:bogol_tsfm}
  \tilde{u}_k(t) &= \alpha_k \, u_k(t) + \beta_k \, u^*_k(t)=\cr
   &=\alpha_k \frac{H^2}{\dot \phi} \frac{1}{\sqrt{2 k^3}}
    (1+ik \eta) e^{-ik \eta}+\cr
   &~~~+\beta_k \frac{H^2}{\dot \phi} \frac{1}{\sqrt{2 k^3}}
    (1-ik \eta) e^{ik \eta}
\end{align}
for slow-roll inflation. If $u_k$ is a vacuum mode function, then $\tilde{u}_k$ corresponds to a mode containing $N_k = \left| \beta_k \right|^2$ particles in mode $k$; also, the Bogoliubov parameters obey $\left| \alpha_k \right|^2 - \left| \beta_k \right|^2 = 1$. 
Only the relative phase $\theta_k$ between $\alpha_k$ and $\beta_k$ is physically relevant so we can write all quantities involving the Bogoliubov parameters in terms of $N_k$ and $\theta_k$ (e.g. $\alpha_k \beta_k^* + \alpha_k^* \beta_k = 2 \sqrt{N_k (N_k + 1)} \cos \theta_k$). $\theta_k$ is an additional unconstrained parameter and for very particular choices it can dramatically enhance $f_{\text{NL}}$; this will be further discussed later.

To use the non-BD $\tilde{u}$, we simply replace $u$ by $\tilde{u}$ in the quantization of $\zeta$ (\ref{eq:qntz_zet}). For the power spectrum, this gives us
\begin{align}\label{eq:nbd_pow_spec}
  P(k)&=|u_k|_{\eta\to0}^2 
    = \frac{H^4}{\dot \phi^2} \frac{1}{2 k^3} \left| \alpha_k + \beta_k \right|^2  \cr
  &= \frac{H^4}{\dot \phi^2} \frac{1}{2 k^3} 
      \left(1 + 2 N_k + 2 \sqrt{N_k (N_k + 1)} \cos \theta_k\right)\;.\cr
\end{align}

While we currently can only conjecture about the exact form of $N_k$, we can place on it certain constraints:
\begin{enumerate}[(a)]
\item The energy content in the fluctuations must be finite. This requires that, in the UV, $N_k =\mathcal{O}(1/k^{4+\delta})$ \cite{[][{. This paper addresses the scalar fluctuations $\varphi$ but for slow-roll inflation, $\zeta=(H/\dot\phi) \varphi$, so the conclusions should still apply.}]Boyanovsky:2006qi} (heuristically at least, this follows from demanding that the energy density $\int d^3k\,k N_k$ converge). Actually, this constraint has only a limited impact on our results; we can only observe a restricted range of wavenumbers so that we are free to posit any form for $N_k$ for high enough wavenumber.
\item\label{item:nk_constr-en} A more restrictive condition comes from demanding that the energy content in the fluctuations be less than that in the unperturbed background field; otherwise, there would be back reaction on the inflaton's dynamics and inflation would not proceed as expected. Roughly, we can write this condition as 
  \begin{align*}
    \int d^3k\,k N_k \lesssim M_\text{pl}^2 H^2.
  \end{align*}
\item The amplitude of the perturbations must be compatible with the observed primordial power spectrum. This constraint is fairly weak since, as we can see from (\ref{eq:nbd_pow_spec}), $P(k) \propto (H^4/\dot \phi^2) (1 + 2 N_k)$ and we cannot independently constrain either the first or second term. 
\item\label{item:running_cond} The running of $N_k$ must conform to the observed spectral tilt $1-n_s\approx 0.032$ \cite{Komatsu:2010fb}. This implies that 
\begin{align*}
  1 - n_s &\equiv -d\log(k^3 \, P(k)\,)/d\log k \cr
   &= -2 \eta_{\text{SR}} + 6 \epsilon - d\log (1+2N_k)/d \log k \cr
   &= 0.032\,.
\end{align*}
Assuming $\epsilon$ and $\eta_{\text{SR}}$ are small, then $N_k$ must change slowly \cite{Agullo:2010ws}.
\end{enumerate}

As in \cite{Holman:2007na}, we will consider $N_k$ with the form 
\begin{align*}
  N_k\approx N_{k,0} e^{-k^2/k_{\text{cut}}^2}\,,
\end{align*}
with some cutoff scale $k_{\text{cut}}$. Condition (\ref{item:nk_constr-en})  gives $N_{k,0} \lesssim M_{\text{pl}}^2 H^2/k_{\text{cut}}^4$. If we take $k_{\text{cut}}\approx\sqrt{M_{\text{pl}} H}$, i.e. the scale of inflation, $N_{k,0}$ can be of order one; for other scenarios, $N_{k,0}$ might be higher or lower. A more careful calculation gives a comparable result \cite{Boyanovsky:2006qi}. If we take the cutoff scale $k_{\text{cut}}$ to be sufficiently larger than the primordial modes observable today then, for observational purposes, $N_k\approx N_{k,0}=\text{const}$; this is one case we consider. We will also calculate the effect of having the cutoff $k_{\text{cut}}$ lie within the observable modes. For some cutoffs, this situation may not satisfy condition (\ref{item:running_cond}) above but it provides an idea about the effect of a varying $N_k$. 

The form of $\theta_k$ is constrained only through the power spectrum (\ref{eq:nbd_pow_spec}), allowing us considerable freedom in its choice. If we think of the non-vacuum initial state as parameterizing unknown physics, then we expect that it is determined by matching the mode function (\ref{eq:bogol_tsfm}) to the initial conditions. It follows that the main (or at least a large) contributor to $\theta_k$ will be the exponential factors $e^{-i k \eta_0}$, $e^{i k \eta_0}$ evaluated at some initial time $\eta_0$; the alternative is that these factors are coincidentally cancelled off in the initial conditions which (since the initial conditions probably represent very different physics) seems unlikely.  Thus, it makes sense to consider $\theta_k \approx k \eta_0$, though we will also explore the effect of a constant $\theta_k$ for very large $k_{\text{cut}}$. 

\begin{widetext}
Returning to the calculation of the bispectrum, we plug (\ref{eq:acn}) and (\ref{eq:qntz_zet}) into (\ref{eq:in-in}) and find that
\begin{align}\label{eq:bispr}
  B_\zeta(k_1, k_2, k_3) 
  = 2 i \frac{\dot \phi^4}{H^6}
    \sum_i \left(\frac{1}{k_i^2}\right) 
        \tilde{u}_{k_1}(\bar{\eta}) \tilde{u}_{k_2}(\bar{\eta}) 
         \tilde{u}_{k_3}(\bar{\eta})
     \int^{\bar{\eta}}_{\eta_0} d\eta \frac{1}{\eta^3} 
       {u'}^*_{k_1} {u'}^*_{k_2} {u'}^*_{k_3} + \cc\;.
\end{align}
We can now replace $u$ by $\tilde{u}$ to account for our initial state, so the integral becomes
\begin{align}\label{eq:nbd-itgtl}
  \int^{\bar{\eta}}_{\eta_0} d\eta \frac{1}{\eta^3} 
    {\tilde{u}}^{'*}_{k_1} {\tilde{u}}^{'*}_{k_2} {\tilde{u}}^{'*}_{k_3}
   = \frac{H^6}{\dot \phi^3} \sqrt{\frac{k_1 k_2 k_3}{8}}
    \int_{\eta_0}^{\bar{\eta}} d\eta \, 
     \left(\alpha_{k_1}^* e^{i k_1 \eta} + \beta_{k_1}^* e^{-i k_1 \eta}\right)
     \left(\alpha_{k_2}^* e^{i k_2 \eta} + \beta_{k_2}^* e^{-i k_2 \eta}\right)
     \left(\alpha_{k_3}^* e^{i k_3 \eta} + \beta_{k_3}^* e^{-i k_3 \eta}\right).
\end{align}
\end{widetext}
For the BD case ($\alpha=1$, $\beta$=0), the integral gives $\left.[1/i(k_1+k_2+k_3)] e^{i(k_1+k_2+k_3)\eta}\right|_{\eta_0}^{\bar{\eta}}$, which in the squeezed limit becomes $\left.(1/2 i\,k_1) e^{2i\, k_1 \eta}\right|_{\eta_0}^{\bar{\eta}}$. If we consider a non-BD initial state, $\beta_k\neq 0$, and we get terms of a different character; for example, there is a $\alpha_{k_1}^* \beta_{k_2}^* \alpha_{k_3}^*$ term which gives $\left.[1/i(k_1-k_2+k_3)] e^{i(k_1-k_2+k_3)\eta}\right|_{\eta_0}^{\bar{\eta}}$, or in the squeezed limit $\left.(1/i\,k_3) e^{i k_3\eta}\right|_{\eta_0}^{\bar{\eta}}$. Thus, \emph{terms like this latter term are enhanced by a factor of $k_1/k_3\gg 1$ in the squeezed limit}.

In principle, it is not clear what happens at the lower limit of integration, when $\eta=\eta_0$. We expect the initial condition for the modes to be set while the modes are within the horizon, i.e. $k \eta_0\gg 1$. Thus, depending on the exact value of $k$ and $\eta_0$, the exponentials will oscillate; when performing our calculations, we will take the average value, i.e. that the exponentials are zero. For the left endpoint, as usual, $\bar{\eta}\to 0$.

\begin{widetext}
In this way, we arrive at an expression for the bispectrum
\begin{align}\label{eq:nbd-bispr}
  B_\zeta(k_1, k_2, k_3) 
  = &\frac{i}{4} \frac{H^6}{\dot \phi^2} \frac{1}{k_1 k_2 k_3}
    \sum_i \left(\frac{1}{k_i^2}\right) 
     \left(\alpha_{k_1} + \beta_{k_1} \right)
     \left(\alpha_{k_2} + \beta_{k_2} \right)
     \left(\alpha_{k_3} + \beta_{k_3} \right)\times\cr
     & \times \int_{\eta_0}^{\bar{\eta}} d\eta \, 
     \left(\alpha_{k_1}^* e^{i k_1 \eta} + \beta_{k_1}^* e^{-i k_1 \eta}\right)
     \left(\alpha_{k_2}^* e^{i k_2 \eta} + \beta_{k_2}^* e^{-i k_2 \eta}\right)
     \left(\alpha_{k_3}^* e^{i k_3 \eta} + \beta_{k_3}^* e^{-i k_3 \eta}\right)+\cc\;,
\end{align}
\end{widetext}
which is written out explicitly in the Appendix. The non-BD part of the shape is primarily squeezed with some enhancement in the folded regime (i.e $k_1 \approx k_2 \approx (1/2) k_3$), coming from the terms mentioned above. As a check on our calculation, note that we recover eq. (4.4) of Maldacena's paper \cite{Malda} if we specialize to the BD case, where $\alpha_k=1$, $\beta_k=0$.

Thus, the non-BD bispectrum is enhanced in the squeezed limit, resulting in a larger $f_{\text{NL}}$.

\subsubsection*{Comparison with previous work}
This result is essentially equivalent to eq. (5.9) of Agullo and Parker \cite{Agullo:2010ws}. However, Agullo and Parker used a density-matrix formalism where the results are essentially expressed in terms the expectation value of products of the occupation number operator $N_{\mbf{k}}$. By contrast, our results are expressed in terms of the Bogoliubov parameters $\alpha_k$, $\beta_k$. While the formalisms are equivalent in principle, the relationship between them is not trivial, since any non-Bunch-Davies state in terms of Bogoliubov parameters is an infinite sum of states with definite occupation numbers \cite{DeWitt:1975ys}. We have not encountered an explicit formula connecting the two descriptions. 

We feel our result is worthwhile because it connects to the formalism of much of the previous literature on the subject. Also, it is often easier to use when trying to hypothesize possible initial states since these are often naturally expressed in Bogoliubov parameters, as we saw in our arguments about the expected behavior of the phase angle $\theta_k$. We do however admit that Agullo's approach has the potential advantage of being able to calculate the bispectrum given any arbitrary initial state, since the space of states expressible in terms of Bogoliubov parameters does not fill the Fock space.

A similar situation was treated in \cite{Holman:2007na} but they were expecting a signal only in the folded limit and did not look for one in the squeezed limit. [Note that there is an error in their eq. (3.16) which further obfuscates the squeezed limit enhancement]. \cite{Meerburg:2009ys} used a similar setup as the present case, examining slow-roll inflation with a modified initial state. Like in the current paper, they performed a 2D projection, though in the flat-sky approximation, which works accurately only for $l\gtrsim 150$ \cite{Fergusson:2008ra} and so is not ideal for calculating the local template (which has a strong signal when one of the $l$'s is small); however, they used the same incorrect bispectrum as \cite{Holman:2007na}, and so their result missed the local limit enhancement. A later paper including two of the same authors \cite{Meerburg:2009fi} found that models with a small speed of sound and a modified initial state can have an enhanced local signal; they also noted that the size of the enhancement depends on the phase angle and they studied the effect of different (constant) phase angles. However, they did not perform the 2D projection so they could not calculate what $f_{\text{NL}}$ would be. Furthermore, the model they used was more complicated, and we feel that are advantages to studying this unexpected behavior in the simple case of standard slow-roll inflation.

\section{The observed value of $f_{\text{NL}}$}
\label{sec:obs_fnl}

Determining the $f_{\text{NL}}$ signal in the CMB requires some additional work. In particular, we must use a transfer function to calculate the present-day temperature anisotropy from the primordial curvature perturbation and then project the 3D CMB onto the 2D projection that we actually observe; this gives us the angular bispectrum. Then, we use the optimal estimator to extract the value of $f_{\text{NL}}$; essentially, this last step involves fitting the measured angular bispectrum to the template angular bispectrum derived from (\ref{eq:loc_bispr}).

The calculations required are described in \cite{Komatsu:2010hc,*Liguori:2010hx}. The angular power spectrum and bispectrum are calculated from their primordial counterparts by 
\begin{gather}
  C_l=\left(\frac{18}{25 \pi}\right) \int dk \, k^2 P_\zeta(k) g_{Tl}^2(k)\,,
\end{gather}
and
\begin{widetext}
\begin{gather}
  \label{eq:redbispr-theory}
  b_{l_1 l_2 l_3} = \left( \frac{6}{5 \pi}  \right)^3
   \int r^2 dr \int dk_1\, dk_2\, dk_3\, (k_1 k_2 k_3)^2 B_\zeta(k_1,k_2,k_3)
    g_{T l_1}(k_1) g_{T l_2}(k_2) g_{T l_3}(k_3) 
    j_{l_1}(k_1 r) j_{l_2}(k_2 r) j_{l_3}(k_3 r)\,,
\end{gather}
\end{widetext}
where $g_{Tl}$ and $j_l$ are the radiation transfer function and spherical Bessel functions, respectively. The angular bispectrum $b_{l_1 l_2 l_3}$ for the non-vacuum initial state is calculated using (\ref{eq:nbd-bispr}) while $b_{l_1 l_2 l_3}^{\text{loc}}$ for the local template bispectrum is calculated using (\ref{eq:loc_bispr}). To calculate $C_l$, one can use (\ref{eq:nbd_pow_spec}). Finally, the optimal estimator for $f_{\text{NL}}$ is given by
\begin{align}\label{eq:fnl_estim}
  f_{\text{NL}} = \frac{1}{N^{\text{loc}}}
   \sum_{l_1 \ge l_2 \ge l_3 \ge 2} I_{l_1 l_2 l_3}^2
     \frac{b_{l_1 l_2 l_3}^{\text{loc}} b_{l_1 l_2 l_3}}
       { \Delta_{l_1 l_2 l_3} C_{l_1} C_{l_2} C_{l_3}},
  \intertext{where}
  \nonumber N^{\text{loc}} \equiv \sum_{l_1 \ge l_2 \ge l_3 \ge 2} I_{l_1 l_2 l_3}^2
  \frac{\left(b_{l_1 l_2 l_3}^{\text{loc}}\right)^2}
    { \Delta_{l_1 l_2 l_3} C_{l_1} C_{l_2} C_{l_3}}\,;
\end{align}
for convenience, we have defined
\begin{align*}
  \Delta_{l_1 l_2 l_3} \equiv
   \begin{cases}
     6 & \text{, if all l's the same}\\
     2 & \text{, if 2 l's the same}\\
     1 & \text{, if all l's different}
   \end{cases},\cr
  I_{l_1 l_2 l_3}\equiv \sqrt{\frac{(2l_1+1)(2l_2+1)(2l_3+1)}{4\pi}}
  \begin{pmatrix}
    l_1 & l_2 & l_3 \\
    0 & 0 & 0
  \end{pmatrix},
\end{align*}
where the matrix above is a Wigner 3-j symbol. 

Given the complicated form of $g_{Tl}$ and of the equations above, analytically solving for $f_{\text{NL}}$ is generally an intractable problem and must be done numerically. Even so, the system is somewhat involved and, in general, presents a difficult computational challenge. This situation is improved if the bispectrum is separable, i.e. if it can be written in the form 
\begin{align*}
  B_\zeta(k_1,k_2,k_3)=\sum_i f_i(k_1) g_i(k_2) h_i(k_3).
\end{align*}
While the local form bispectrum (\ref{eq:loc_bispr}) is of this form, the non-vacuum slow-roll bispectrum (\ref{eq:nbd-bispr}) is not. However, since calculating $f_{\text{NL}}$ requires fitting a bispectrum to the local form, the result is primarily dependent on the regime where the local form peeks, i.e. the squeezed limit $k_3 \ll k_1 \approx k_2$. More formally, one may examine (\ref{eq:redbispr-theory}), (\ref{eq:fnl_estim}), -- noting that $g_{Tl}(k)$ peaks when $k\approx l / r_{\text{LS}}$, where $r_{\text{LS}}$ is the comoving distance to the surface of last scattering -- and arrive at the same conclusion. In other words, we can make our bispectrum (\ref{eq:nbd-bispr}) separable by assuming that it is evaluated in the local limit, where $k_3\ll k_1$. Then, our bispectrum becomes:
\begin{gather}\label{eq:bispr_low_order}
  B_\zeta(k_1,k_2,k_3) =
    \frac{H^6}{\dot\phi^2} \frac{1}{4 k_1 k_2 k_3^4} \mathcal{N},
\end{gather}
where
\begin{align}\label{eq:nbd-bispr_low_order}
  \mathcal{N} &\equiv\prod_i(\alpha_{k_i}+\beta_{k_i}) \;
     (\alpha_{k_1}^* \beta_{k_2}^* \alpha_{k_3}^* 
       + \beta_{k_1}^* \alpha_{k_2}^* \alpha_{k_3}^*) -\cr
    &\phantom{=}~~~~~-(\alpha \leftrightarrow \beta) + \cc =\cr
   &= 2 N_1 + 2 N_2 + 4 N_1 N_2 \cr
    &\phantom{=}~+ 4 N_1 \sqrt{N_2 (N_2 + 1)} \cos \theta_2 +\cr
    &\phantom{=}~+ 4 N_2 \sqrt{N_1 (N_1 + 1)} \cos \theta_1 +\cr
    &\phantom{=}~+  2 \sqrt{N_1 (N_1 + 1)} \cos \theta_1  
     + 2 \sqrt{N_2 (N_2 + 1)} \cos \theta_2 +\cr
    &\phantom{=}~+  4 \sqrt{N_1 N_2 (N_1 + 1) (N_2 + 1)} 
      \cos \theta_1  \cos \theta_2 +\cr
    &\phantom{=}~+ 4 \sqrt{N_1 N_2 (N_1 + 1) (N_2 + 1)} 
      \sin \theta_1 \sin \theta_2 +\cr
    &\phantom{=}~+ 4 \sqrt{N_1 N_3 (N_1 + 1) (N_3 + 1)} 
      \sin \theta_1 \sin \theta_3 +\cr   
    &\phantom{=}~+ 4 \sqrt{N_2 N_3 (N_2 + 1) (N_3 + 1)} 
      \sin \theta_2 \sin \theta_3\;,
\end{align}
where $N_i \equiv N_{k_i}, \theta_i \equiv \theta_{k_i}$.
(To ascertain the reasonableness of this approximation, we performed the calculation to the next order in $k_3$ and found results that were 100 times smaller).  Note that the standard $1-n_s$ terms are absent because they are lower order in $k_3$.

To find $f_{\text{NL}}$, we plugged (\ref{eq:bispr_low_order}) into (\ref{eq:redbispr-theory}) and (\ref{eq:fnl_estim}) and numerically performed the integrals in (\ref{eq:redbispr-theory}). To determine the radiation transfer function $g_{Tl}$, we used the gtFAST software \cite{gtfast}, based on CMBFAST 4.0 \cite{Seljak:1996is}; for input, we assumed $h=70.3$, $T_{\text{CMB}}=2.725\text{ K}$, $\Omega_b=0.0451$, $\Omega_c=0.226$, $\Omega_\Lambda=0.729$, $\Omega_\nu=0$, $N_\nu=3.04$, $\tau=0.088$, and $z_{\text{reion}}=10.4$ (from \cite{Komatsu:2010fb}). We summed $l$'s up to 2000, to reflect the signal available in the CMB temperature power spectrum.

\section{Results}
\label{sec:results}

As described in \S~\ref{sec:swr-inf}, we calculated $f_{\text{NL}}$ for various values of the cutoff $k_{\text{cut}}$:
\\

\noindent\textbf{a) $\mathbf{k_{\text{cut}}}$ is very large, so that, $\mathbf{N_k\approx N_{k,0}\approx\text{const}}$:} \\
We supposed either that $\theta_k\approx\text{const}$ or $\theta_k \approx k \eta_0$. For the latter case, the terms in (\ref{eq:nbd-bispr_low_order}) with one trigonometric function will clearly average out to zero in (\ref{eq:redbispr-theory}). The terms with two trigonometric functions initially appear as if they might contribute, since $k_1 \approx k_2$, giving e.g. $\cos \theta_1 \cos \theta_2 \approx \cos^2 \theta_1$. However, the integrand in (\ref{eq:redbispr-theory}) is sensitive only to $k$ values that satisfy a triangle inequality (while this is not obvious from (\ref{eq:redbispr-theory}), it is clear from the derivation of that equation) so that, for example, $k_2 \in [k_1 - k_3, k_1 + k_3]$. Even though $k_3\ll k_1 \approx k_2$, $\eta_0$ is sufficiently large ($k_3 \eta_0 \gg 0$ because we suppose that the initial conditions are set when $k_3$ is within the horizon) that these terms will oscillate greatly during the integral over $k_2$ and average out. Thus, we will be left with only the first line of the second equality in (\ref{eq:nbd-bispr_low_order}): $\mathcal{N}_{\theta_k \approx k \eta_0} = 4 (N_{k,0} + N_{k,0}^2)$. 

For either form of $\theta_k$, the numerical integral is the same. Performing the integral, we found
\begin{align}\label{eq:nk_const_result_general}
  f_{\text{NL}} = 76 \frac{\dot \phi^2}{H^2} 
    \frac{\mathcal{N}}{\tilde{P}^2}
   = 1.52 \,\frac{\epsilon}{0.01}
    \frac{\mathcal{N}}{\tilde{P}^2},
\end{align}
where we have defined $P(k) = H^4 / (2 k^3 \dot \phi^2)\, \tilde{P}$, so that $\tilde{P}(k)$ equals the term in parenthesis in (\ref{eq:nbd_pow_spec}). We have expressed the answer in terms of the slow-roll parameter $\epsilon$. If the single-field relation $r=16\epsilon$ still holds for this situation (where $r$ is the tensor-scalar ratio), the bound $r < 0.2$ \cite{Komatsu:2010fb} yields $\epsilon < 0.013$; however, since we have a non-vacuum initial state, this relation may no longer hold, and $\epsilon$ could be larger than otherwise expected (but the slow-roll approximation breaks down for $\epsilon \gtrsim 0.1)$.

For the case that $\theta_k \approx k \eta_0$, (\ref{eq:nk_const_result_general}) becomes
\begin{align}\label{eq:nk_const_no_theta_result}
  f_{\text{NL}} = 1.52 \,\frac{\epsilon}{0.01}
   \left( 1 - \frac{1}{1 + 4N_{k,0} + 4N_{k,0}^2}\right).
\end{align}
We thus have $f_{\text{NL}}\lesssim 1.6 (\epsilon/0.01)$; for the standard bound on $\epsilon$ this is indistinguishable in the CMB -- since the best the CMB can do is $\Delta f_{\text{NL}}\approx 3$ \cite{Baumann:2008aq}. For larger values of $\epsilon$, this might actually be observable by future CMB satellites.

\begin{figure}
  \centering
  \scalebox{0.58}{\includegraphics{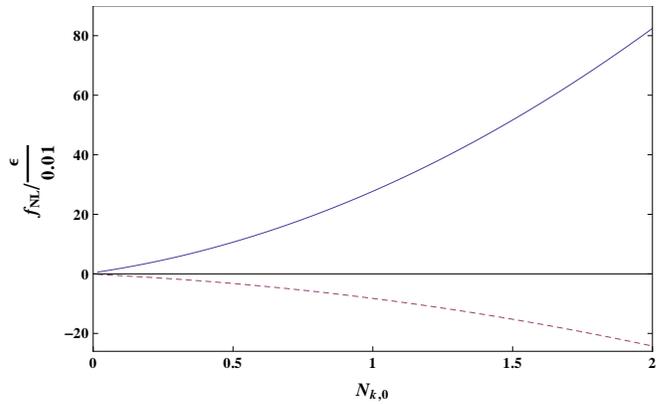}}
  \caption{For very large $k_{\text{cut}}$, where $N_k \approx N_{k,0} = \text{const}$, and fine-tuned values of $\theta_k=\text{const}$, $f_{\text{NL}}$ can be large. The maximum values (upper solid line) are achieved for $\cos \theta = (1 - 12 N - 12 N^2) / 6 \sqrt{N (N+1)} (1+2N)$, and the minimum values (lower dashed line) are achieved for $\cos \theta=-1$.}
  \label{fig:fnl-for-thet-const}
\end{figure}

If we instead consider that $\theta_k\approx\text{const}$, we can get significantly larger results and even negative $f_{\text{NL}}$. Indeed, finding the extrema of (\ref{eq:nk_const_result_general}) with respect to a constant $\theta_k$ for $N_{k,0} > 0.017$, one finds a maximum of 
\begin{align}\label{eq:nk,theta_const_max}
  f_{\text{NL}} = 1.52 \,\frac{\epsilon}{0.01} 
    \left(\frac{1}{4}+ 9 N_{k,0} + 9 N_{k,0}^2\right)
\end{align}
and a minimum of 
\begin{align}\label{eq:nk,theta_const_min}
  f_{\text{NL}} = - 1.52 \,\frac{\epsilon}{0.01} 
    \Big[&2 N_{k,0} (1 + N_{k,0}) +\cr
   &+ \sqrt{N_{k,0} (1+N_{k,0})} (1+ 2 N_{k,0})\Big],\cr
\end{align}
where $N\equiv N_{k,0}$. These values are depicted in Fig. \ref{fig:fnl-for-thet-const} as a function of $N_{k,0}$.
\\

\noindent\textbf{b) $\mathbf{k_{\text{cut}}}$ lies within the observable modes:} \\
As acknowledged earlier, if $k_{\text{cut}}$ lies within the observable modes, there might be a conflict with measured power spectrum results; however, because of the importance of fully investigating possible inflationary scenarios and understanding their consequences, we felt it was worthwhile to perform the calculation.  Using a simple formula, one can relate the CMB-anisotropy multipole $l$ to the wavenumber $k$ most responsible for producing it at the surface of last scattering: $k\approx l/r_{\text{LS}}$ \footnote{This result can be seen by looking at the transfer function, as noted in \S~\ref{sec:obs_fnl}. Heuristically, it can also be derived by noting that, for small angles, $l\approx \pi/\theta$, where $\theta$ is the angular size of the corresponding fluctuations; then, using simple trigonometry, $r_{\text{LS}} \theta \approx \lambda/2 = \pi/k$, yielding the relation.}, where $r_{\text{LS}}=14400\text{ Mpc}^{-1}$ is the comoving distance to the surface of last scattering. We considered values of $k_{\text{cut}}$ corresponding to a cutoff at $l = 500$ and $l = 1500$. We further calculated for $\theta_k\approx k \eta_0$, as well as for $\theta_k \approx k/k^*$ where $k^*\ll k_3$ so that the terms with trigonometric functions in (\ref{eq:nbd-bispr_low_order}) average out as in (a). The results for several values of $N_{k,0}$ are listed in Table \ref{tab:non-const-nk}. We see that the values for $f_{\text{NL}}$ are much smaller than for $N_k\approx\text{const}$.
\begin{table}
  \begin{tabular}{cccc}
    \hline
    \noalign{\smallskip}
    \centering $l_{\text{cut}}$ & $k_{\text{cut}}$ & $N_{k,0}$ &
    $\displaystyle f_{\text{NL}} \left( \frac{\epsilon}{0.01} \right)^{-1}$\\
    \hline 
    \noalign{\smallskip}
    very large
     & very large
     & \ \ $N_{k,0}$\ \
     & $1.52
        [ 1 - (1 + 4N_{k,0} + 4N_{k,0}^2)^{-1} ]$ \\
    \hline
    \centering\multirow{3}{*}{1500} 
     & \centering\multirow{3}{*}{$0.104\text{ Mpc}^{-1}$}
     &0.1 & 0.14\\
     && 1 & 0.44\\
     && 10 & 0.46\\
    \hline
    \noalign{\smallskip}
    \centering\multirow{3}{*}{500} 
     & \centering\multirow{3}{*}{$0.0347\text{ Mpc}^{-1}$} 
     & 0.1 & $5.4\times10^{-3}$\\
     && 1 & $1.9\times10^{-2}$\\
     && 10 & $2.32\times10^{-2}$\\
    \hline
  \end{tabular}
  \caption{The observed $f_{\text{NL}}$ for $N_k=N_{k,0} e^{-k^2/k_{\text{cut}}^2}$. The top line shows the result for $k_{\text{cut}}$ very large, so that $N_k\approx\text{const}$. For cutoff $k_{\text{cut}}\equiv l_{\text{cut}}/r_{\text{LS}}$, we get a suppression of $N_k$ for the wavenumbers that contribute primarily to multipoles larger than $l_{\text{cut}}$ in the CMB temperature anisotropy.}
  \label{tab:non-const-nk}
\end{table}

\section{Discussion and conclusion}
\label{sec:discuss}

The conclusions from this work depend on the form of $\theta_k$ and the size of $\epsilon$. In the most conservative case, we suppose that  $\theta_k\approx \eta_0 k$ and $\epsilon\lesssim 0.01$, and we find from (\ref{eq:nk_const_no_theta_result}) that $f_{\text{NL}} \lesssim 3$ in the CMB and thus undetectable. This case conforms to the general idea that a detection of $f_{\text{NL}}$ would disfavor single-field inflation. However, if we allow either $\theta_k$ to be constant or $\epsilon$ to be slightly elevated (which, as mentioned in \S~\ref{sec:results}  is possible for a non-Bunch-Davies initial state), we can get a larger $f_{\text{NL}}$ that could be measured. The second of these scenarios -- a larger $\epsilon$ -- can produce a (barely) measurable $f_{\text{NL}}\approx 15$ if $\epsilon\approx 0.1$ and $N_k\gtrsim 1$. The first scenario, where $\theta_k\approx\text{const}$, can produce a much larger and/or negative $f_{\text{NL}}$, depending on the value of $\theta_k$, as we see from (\ref{eq:nk,theta_const_max}), (\ref{eq:nk,theta_const_min}), and Fig. \ref{fig:fnl-for-thet-const}, rising to $\approx 30$ for $N_k=\mathcal{O}(1)$ and $\epsilon\approx 0.01$. Thus, a detection of local-type $f_{\text{NL}}$ alone would not rule out single-find inflation.

It might seem curious that $f_{\text{NL}}$ in (\ref{eq:nk_const_no_theta_result}) does not blow up for very large $N_{k,0}$. To understand this, note that (\ref{eq:loc_bispr_sq_lim}) shows that $f_{\text{NL}}$ roughly measures the size of the bispectrum in the squeezed limit \emph{relative to the product of power spectra $P(k_1) P(k_3)$}. However, for large $N_{k,0}$, $P(k)\propto N_{k,0}$ and $B_\zeta\propto N_{k,0}^2$, so that $B_\zeta$ increases exactly like $P(k_1) P(k_3)$ with respect to $N_{k,0}$, and $f_{\text{NL}}$ is not enhanced. In fact, the non-numerical factors in (\ref{eq:nk_const_no_theta_result}) can be derived by supposing
\begin{align*}
  f_{\text{NL}} \approx \frac{B_\zeta}{P(k_1) P(k_3)}.
\end{align*}

In examining the effect of $N_{k,0}$ of $f_{\text{NL}}$ in Table \ref{tab:non-const-nk}, we see that $f_{\text{NL}}$ decreases for decreasing $l_\text{cut}$. This is consistent with expectations since the bispectrum we found was enhanced by $k_1 / k_3$, and a lower $l_\text{cut}$ prevents the contribution of modes where $k_1$ is large. Along the same lines, one might also wonder about the effect of other forms of $N_k$, for example a sharp peak at some momentum $k$ due to a condensate or particle production. The effect of such a form would likely depend on the specifics of the model under consideration. If the boosted power is at small or large scales, where $k_1 / k_3$ is large, it will increase the observability of the $f_{\text{NL}}$ enhancement; however, the observability is also a function of the width of the peak, since an overly narrow peak will be washed out as the bispectrum is projected to 2D and then used to calculate $f_{\text{NL}}$.

As a last note, we remark on the fact that our results produce $f_{\text{NL}}$ larger than $(5/12) (1-n_s)\approx 0.01$, as predicted by the single-field consistency relation. To reconcile this apparent discrepancy, we examine the integrals which give the enhanced terms, as described below (\ref{eq:nbd-itgtl}). These integrals yield terms like $\left.e^{i k_3 \eta}\right|_{\eta_0}^\eta$. In the exact squeezed-limit of the consistency relation, $k_3$ is taken to be arbitrarily small, so $|k_3 \eta_0| \ll 1$ and these terms are zero; the only pieces remaining reproduce the Maldacena squeezed-limit result. This explanation is somewhat unsatisfying because $|k_3 \eta_0| \ll 1$ implies that the initial condition was set when $k_3$ was far outside the horizon; however, one may argue that unknown, early universe effects were responsible for this and, in any case, the consistency relation does hold. Thus, we see that the consistency relation is a powerful tool but it only applies perfectly in a certain limit (a similar conclusion was reached in \cite{Ganc:2010ff}).

\begin{acknowledgments}
We particularly want to thank Eiichiro Komatsu for his guidance and advice. Ivan Agullo provided many helpful comments and ideas, as well as noting several errors. Xingang Chen made several useful suggestions. Also, we extend appreciation to Willy Fischler and Sonia Paban for helping us understand the conditions for a viable initial state. This work is supported in part by NSF grant PHY-0758153.
\end{acknowledgments}

\appendix*

\section{Explicit Bispectrum}\label{app:full-bispectrum}

From (\ref{eq:nbd-bispr}), we easily write out the bispectrum explicitly:

\begin{widetext}
\begin{align}\label{eq:explicit-bispr}
  B_\zeta&(k_1, k_2, k_3)=\cr
   &\begin{aligned}= &\frac{i}{4} \frac{H^6}{\dot \phi^2} \frac{1}{k_1 k_2 k_3}
     \sum_i \left(\frac{1}{k_i^2}\right) 
      \left(\alpha_{k_1} + \beta_{k_1} \right)
      \left(\alpha_{k_2} + \beta_{k_2} \right)
      \left(\alpha_{k_3} + \beta_{k_3} \right)\times\cr
    & \times \int_{\eta_0}^{\bar{\eta}} d\eta \, 
      \left(\alpha_{k_1}^* e^{i k_1 \eta} + \beta_{k_1}^* e^{-i k_1 \eta}\right)
      \left(\alpha_{k_2}^* e^{i k_2 \eta} + \beta_{k_2}^* e^{-i k_2 \eta}\right)
      \left(\alpha_{k_3}^* e^{i k_3 \eta} + \beta_{k_3}^* e^{-i k_3 \eta}\right)+\cc\approx
    \end{aligned}\cr
   &\begin{aligned}\approx &\frac{1}{4} \frac{H^6}{\dot \phi^2} \frac{1}{k_1 k_2 k_3}
     \sum_i \left(\frac{1}{k_i^2}\right) 
      \left(\alpha_{k_1} + \beta_{k_1} \right)
      \left(\alpha_{k_2} + \beta_{k_2} \right)
      \left(\alpha_{k_3} + \beta_{k_3} \right)\times\cr
    &\begin{aligned} \times 
      \Big[&\frac{1}{k_1 + k_2 + k_3} 
         (\alpha_{k_1}^* \alpha_{k_2}^* \alpha_{k_3}^* - \beta_{k_1}^* \beta_{k_2}^* \beta_{k_3}^*)
        +\frac{1}{k_1 + k_2 - k_3}
         (\alpha_{k_1}^* \alpha_{k_2}^* \beta_{k_3}^* - \beta_{k_1}^* \beta_{k_2}^* \alpha_{k_3}^*)+\cr
       &+\frac{1}{k_1 - k_2 + k_3} 
         (\alpha_{k_1}^* \beta_{k_2}^* \alpha_{k_3}^* - \beta_{k_1}^* \alpha_{k_2}^* \beta_{k_3}^*)
        +\frac{1}{k_1 - k_2 - k_3}
         (\alpha_{k_1}^* \beta_{k_2}^* \beta_{k_3}^* - \beta_{k_1}^* \alpha_{k_2}^* \alpha_{k_3}^*)
        \Big]
    \end{aligned}\cr
   &+\cc\;,\end{aligned}\cr
\end{align}
\end{widetext}
where, in the last equation we let $\bar{\eta}\to0$ and averaged over $\eta_0$, as discussed in \S~\ref{sec:swr-inf}.

\bibliography{paper-bib}

\end{document}